\documentclass[letterpaper, 10 pt, conference]{ieeeconf}  

\IEEEoverridecommandlockouts                              

\title{\LARGE \bf
	Data-Driven Distributionally Robust Safety Verification Using Barrier Certificates and Conditional Mean Embeddings
}

\author{Oliver Sch\"on$^{1}$, Zhengang Zhong$^{2}$, and Sadegh Soudjani$^{3}$
	\thanks{This work is supported by the following grants: EPSRC EP/V043676/1, EIC 101070802, and ERC 101089047.}
	\thanks{$^{1}$Oliver Sch\"on is with the School of Computing, Newcastle University, Newcastle upon Tyne, NE4 5TG, United Kingdom.
		{\tt\small o.schoen2@ncl.ac.uk}}%
	\thanks{$^{2}$Zhengang Zhong is with the Sargent Centre for Process Systems Engineering, Imperial College London, London, United Kingdom.
		{\tt\small z.zhong20@imperial.ac.uk}}%
	\thanks{$^{3}$Sadegh Soudjani is with the Max Planck Institute for Software Systems, Kaiserslautern, Germany.
		{\tt\small sadegh@mpi-sws.org}}%
}

\usepackage{cite}
\usepackage{amsmath,amssymb,amsfonts,amsthm}
\usepackage{algorithmic}
\usepackage{graphicx}
\usepackage{textcomp}
\usepackage{xcolor}
\def\BibTeX{{\rm B\kern-.05em{\sc i\kern-.025em b}\kern-.08em
		T\kern-.1667em\lower.7ex\hbox{E}\kern-.125emX}}
\usepackage{tikz}
\usepackage{comment}

\newtheoremstyle{theoremdd}
{\topsep}
{\topsep}
{\itshape}
{0pt}
{\bfseries}
{:}
{ }
{\thmname{#1}\thmnumber{ #2}\boldmath\textbf{\thmnote{ (#3)}}} 
\theoremstyle{theoremdd}
\makeatletter 
\renewenvironment{proof}[1][\proofname]{\par
	\pushQED{\qed}%
	\normalfont \topsep6\p@\@plus6\p@\relax
	\trivlist
	\item\relax
	{\itshape
		#1\@addpunct{:}}\hspace\labelsep\ignorespaces
}{%
	\popQED\endtrivlist\@endpefalse
}
\makeatother

\newtheorem{problem}{Problem}
\newtheorem{definition}{Definition}
\newtheorem{theorem}{Theorem}
\newtheorem{proposition}{Proposition}
\usepackage[framemethod=TikZ]{mdframed}
\usetikzlibrary{shadows}
\definecolor{grayfilling}{gray}{0.95} 
\definecolor{grayshadow}{gray}{0.5} 
\newenvironment{prob}
{\begin{mdframed}[backgroundcolor=grayfilling, shadow=true, shadowsize=5.5pt, shadowcolor=grayshadow]
		\begin{problem}}
		{\end{problem}\vspace{.4em}\end{mdframed}}
\newtheorem{assumption}{Assumption}       
\newtheorem{remark}{Remark}

\newcommand{\ambRKHS}{\mathcal{C}} 
\newcommand{\B}{B}
\newcommand{\borel}[1]{\mathcal{B}(#1)}

\newcommand{\cdotx}{\,\cdot\,}

\newcommand{\E}{\mathbb{E}}
\newcommand{\Hilbert}{\mathcal{H}}
\newcommand{\innerH}[3]{\langle #1, #2 \rangle_{#3}} 
\newcommand{\M}{\mathbf{M}}
\newcommand{\N}{\mathbb{N}}
\newcommand{\norm}[1]{\left|\left|#1\right|\right|}
\renewcommand{\P}{\mathbb{P}}
\newcommand{\R}{\mathbb{R}}
\newcommand{\U}{\mathbb{U}}
\newcommand{\satisfies}{\vDash}
\newcommand{\T}{^\top}
\newcommand{\Tr}{\mathbf{t}}
\newcommand{\p}{\mathbf{p}}
\newcommand{\X}{\mathbb{X}}

\newcommand{\Y}{\mathbb{Y}}

\DeclareMathAlphabet{\mathmybb}{U}{bbold}{m}{n} 
\newcommand{\1}{\mathmybb{1}}
\usepackage{upgreek}

\usepackage[author={Oliver}]{pdfcomment}
\pdfcommentsetup{color=cyan!30}
\newcommand{\new}[1]{\textcolor{blue}{#1}}
\newcommand{\gray}[1]{\textcolor{gray}{#1}}
\colorlet{cherryred}{red!80!black}
\newcommand{\red}[1]{\textcolor{red}{#1}}

\newcommand{\Oliver}[1]{{\color{green!70!black} [Oliver]: #1}}
\usepackage{hyperref} 
\hypersetup{
	colorlinks   = true, 
	urlcolor     = cyan, 
	linkcolor    = cyan, 
	citecolor    = cyan 
}

\graphicspath{{./Figures/}} 
\usepackage[outdir=./]{epstopdf}

\begin{document}

\maketitle
\thispagestyle{empty}
\pagestyle{empty}

\begin{abstract}
Algorithmic verification of realistic systems to satisfy safety and other temporal requirements has suffered from poor scalability of the employed formal approaches. 
To design systems with rigorous guarantees, many approaches still rely on exact models of the underlying systems.
Since this assumption can rarely be met in practice, models have to be inferred from measurement data or are bypassed completely.
Whilst former usually requires the model structure to be known a-priori and immense amounts of data to be available, latter gives rise to a plethora of restrictive mathematical assumptions about the unknown dynamics.
In a pursuit of developing scalable formal verification algorithms without shifting the problem to unrealistic assumptions, we employ the concept of barrier certificates, which can guarantee safety of the system, and learn the certificate directly from a compact set of system trajectories. We use conditional mean embeddings to embed data from the system into a reproducing kernel Hilbert space (RKHS) and construct an RKHS ambiguity set that can be inflated to robustify the result w.r.t. a set of plausible transition kernels.
We show how to solve the resulting program efficiently using sum-of-squares optimization and a Gaussian process envelope.
Our approach lifts the need for restrictive assumptions on the system dynamics and uncertainty, and suggests an improvement in
the sample complexity of verifying the safety of a system on a tested case study compared to a state-of-the-art
approach.
\end{abstract}


\section{INTRODUCTION}
Modern engineering is facing many challenges to provide us with products exhibiting some form of ``intelligent'' behavior and at the same time meeting a wealth of regulations. For example, an autonomous car trained on a finite set of scenarios is expected to perform better than human drivers whilst operating safely even in unprecedented scenarios. Failure to do so is likely to result in catastrophic events, e.g., autonomous cars blocking the road and delaying patient care. 

To provide control design techniques for systems in uncertain environments to exhibit the required (safety) behavior \emph{with guarantees}, formal methods for verification and controller synthesis are progressively gaining attention \cite{BK08,belta2017formal,Lavaei_Survey,yin2024formal}.
Despite the maturity of this research line, the available approaches are still at the early stage of being applied in the engineering of realistic systems.
Most approaches rely on exact models of the underlying system dynamics, which are in general hard or impossible to obtain.
In an effort to address these shortcomings, data-driven methods are being exploited for their ability to infer models from data. 
However, to identify models or reason on the samples directly, prior knowledge about the model structure and dynamics must be available, which makes the results applicable to only selected classes of systems, such as linear or control affine systems~\cite{yin2024formal,Lavaei_Survey}.
Furthermore, available data-driven approaches require impracticable amounts of data to establish any useful safety guarantees (e.g., billions of samples for a 3D system~\cite{Salamati2021DDCBC}).

Recent advances in the understanding of modern machine learning algorithms have sparked a deeper interest in the representation of models as functions living in so-called \emph{reproducing kernel Hilbert spaces} (RKHS) \cite{Berlinet2004RKHSProbStat, Steinwart2006SVM, scholkopf2002learning}. Not only has this enabled a deeper understanding of the inner workings of complex functional methods, but it has led to the development of mathematical concepts such as \emph{conditional mean embeddings} (CME)
\cite{Klebanov2020RigorousCME,Song2009CondEmbed}, which allow us to compute the conditional expectation of a nonlinear function of a random variable as an inner product. Furthermore, CMEs allow us to approximate the behavior of stochastic processes based on finite data sets. This renders CMEs a powerful tool for computing the expected behavior of stochastic systems.
Paired with statistical results based on concentration inequalities, probabilistic guarantees for the latent data-generating system can be obtained from small and noisy data sets~\cite{li2022optimal}.
With a focus on control problems, optimization via kernel-based tools such as CMEs holds a range of attractive benefits over other means of data-driven optimization. The most prominent advantages include 
translating general nonlinear programs to linear programs in RKHSs;
bypassing intermediate problems such as density estimation, Lipschitz constants, and numerical integration;
and facilitating the direct computation of conditional expectations on the available data samples.
Incidentally, existing approaches for data-driven formal verification and synthesis are limited by the same challenges whilst CMEs can enable the approximate reformulation of the original problem using only a finite number of observed transitions of the original system without imposing prior assumptions on the structure of the system dynamics or uncertainty.

In this work, we utilize state-of-the-art CME theory to verify safety of stochastic systems without model knowledge via the concept of barrier certificates (BC) \cite{prajna2006barrier}. We reformulate the probabilistic BC constraints into a data-driven optimization problem and obtain distributionally robust results by constructing an ambiguity set of candidate transition kernels.
We solve the resulting problem using sum-of-squares (SOS) optimization \cite{Parrilo2003SOS} and compute a Gaussian process envelope with a bounded approximation error.
Our approach relaxes the restrictive assumptions of prior approaches and suggests an improvement in the sample complexity on a tested case study in comparison with an alternative approach \cite{Salamati2021DDCBC}.

\smallskip

The rest of the paper is organized as follows.
After a brief review of related work, we present the preliminaries and problem statement in Sec.~\ref{sec:preAndProblem}.
The concepts of RKHS and kernel mean embeddings for (conditional) distributions are introduced in Sec.~\ref{sec:RKHS}.
We derive inequality constraints for computing BCs directly from data in Sec.~\ref{sec:ddcbc} and provide an SOS-based solution in Sec.~\ref{sec:sos},
followed-up by a case study and concluding remarks in Secs.~\ref{sec:numerical_studies} and \ref{sec:conclusion}, respectively.

\smallskip

\noindent\textbf{Related Work:}
Many
data-driven approaches have been developed for verification and control synthesis of uncertain systems with control-affine dynamics~\cite{Cohen2022, Lopez2022uCBF, Jagtap2020CBCGP}.
Recent data-driven abstraction-based approaches have also been proposed \cite{jackson2020safety, kazemi2022datadriven,Chekan2023UncertainConstraints,schoen2023bayesian,zhang2024formal, makdesi2023data, gracia2023distributionally}. 
For discrete-time systems, data-driven computation of barrier certificates has been studied in \cite{Salamati2021DDCBC} with an approach that has exponential sample complexity.
Temporal logic control via model-free reinforcement learning is studied in the papers \cite{kazemi2020fullLTL,lavaei2020RL}.
Gaussian processes are used by the authors of \cite{Jagtap2020CBCGP,Wang2018CBF} to learn partially unknown dynamics of nonlinear systems whilst assuming the affine control-dependent part of the dynamics to be known.  
The work \cite{Wang2018CBF} utilizes barrier functions for safe exploration of the state space. Similarly, the work \cite{lindemann2021learning} considers systems with known affine hybrid dynamics and learns a safe state set via control barrier functions from expert demonstrations.
CMEs have been used recently in a correct-by-design approach \cite{Romao2023DRControl}, which embeds the transition kernel for abstraction-based control based on value iteration. In contrast, we consider CMEs in an abstraction-free setting.

\section{PRELIMINARIES AND PROBLEM STATEMENT}\label{sec:preAndProblem}
\noindent\textbf{Notation:}
We denote the sets of positive integers and non-negative reals as $\N_{>0}$ and $\R_{\geq 0}$, respectively.
Consider a Polish sample space $\X$ \cite{bogachev2007measure}.
Let $(\X,\borel{\X},\P)$ be the underlying probability space equipped with a Borel $\sigma$-algebra $\borel{\X}$ defined over $\X$, i.e., the smallest $\sigma$-algebra containing open subsets of $\X$, and a probability measure $\P$. 
	For a random variable $X$, let
	$p_X$ be 
	the pushforward probability measure of $\P$ under $X$ such that $X\sim p_X(\cdotx)$.
The expected value of a function $f(X)$ on $\X$ is written as $\E_{p_X}[f(X)]$. If	clear from the context, we abbreviate and write $\E[f(X)]$.
%
We denote the set of all probability measures for a given measurable space $(\X,\borel{\X})$ as $\mathcal{P}(\X)$.
For two measurable spaces $(\X,\borel{\X})$ and $(\Y,\borel{\Y})$, a \emph{probability kernel} is a mapping $\p: \X \times \borel{\Y}\rightarrow  [0,1]$ such that $\p(X=x,\cdotx):\borel{\Y}\rightarrow[0,1]$ is a probability measure for all $x\in\X$, and $\p(\cdotx, B): \X\rightarrow [0,1]$ is measurable for all  $B\in\borel{\Y}$.
A probability kernel associates to each point $x\in\X$ a measure denoted by $\p(\cdotx|X=x)$.
The \emph{Dirac delta} measure $\delta_a:\borel{\X}\rightarrow [0,1]$ concentrated at a point $a\in\X$ is defined as $\delta_a(A)=1$ if $a\in A$ and $\delta_a(A)=0$ otherwise, for any measurable set $A\in\borel{\X}$.
We denote the uniform distribution over $\X$ as $\mathcal{U}_\X$ with realizations $X\sim\mathcal{U}_\X(\cdotx)$.

%

The transpose of a vector or matrix $A$ is indicated as $A\T$. 
Let the $N\times N$ dimensional identity matrix be given by $I_N$.
For $\X$, let $X_N:=[x_i]_{i=1}^N$ be a column vector with $x_i\in\X$.
We denote the element-wise evaluation of a function $f:\X\rightarrow\R$ on $X_N$ as $f(X_N):=[f(x_i)]_{i=1}^N$.
Similarly, we may write $A = [a_{ij}]_{i,j=1}^N$ to denote a matrix with its elements.


\subsection{Discrete-time stochastic systems}
In this work, we consider systems expressible as Markov chains over continuous spaces, formally defined as follows.
\begin{definition}[Markov chain (MC)]
	An MC is a tuple $\M=(\X,\X_0,\Tr)$, comprising
	a state space $\X$ with states $x\in\X$;
	initial states $x_0\in \X_0\subset\X$;
	and a probability kernel $\Tr:\X\times\borel{\X}\rightarrow[0,1]$.
\end{definition}
In every execution, given a current state $x\in\X$,
 the consecutive state $x_{+}\in\X$ of the MC is obtained as a realization $x_{+}\sim\Tr(\,\cdot\,|X=x)$.
As one particular example of a class of systems that can be captured using MCs, we may consider, e.g., dynamical systems with discrete-time stochastic dynamics, i.e., systems of the form
\begin{equation}
	\label{eq:model}
	\M: \left\{ \begin{array}{ll}
		x_{t+1}&= f(x_t,w_t),\quad w_t\sim p_w(\cdotx),\\
		y_t& = x_t,
	\end{array} \right.
\end{equation}
where the system state and observation at the $t^{\text{th}}$ time-step are denoted by $x_t$ and $y_t$, respectively. 
The state evolution of the system, described by the function $f:\X\times\mathbb{W}\rightarrow\X$,
is subject to {independent, identically distributed (i.i.d.) noise} $w_t\sim p_w(\cdotx)$ supported on a set $\mathbb{W}$.


\subsection{System safety}
In this paper, we wish to verify safety of a system.
%
For an MC $\M$, we characterize \emph{safety} as a property of state. More precisely, a system $\M$ is said to be safe if for any initial state $x_0\in \X_0$ the system does not enter an \emph{unsafe set} of states $\X_u\subset\X$.
Note that stochastic systems in general do not admit a binary notion of safety. 
Therefore, we quantify the \emph{probability} of reaching $\X_u$ in a given finite horizon $T\in\N_{>0}$.
A \emph{safety specification} $\psi:=(\X_u,T)$ is hence fully characterized by an unsafe set $\X_u$ and a time horizon $T$.
The system $\M$ satisfies $\psi$ with probability at least $p_\psi\in[0,1]$ if the probability of its trajectories initialized in $\X_0$ and not entering $\X_u$ within horizon $T$ is greater than or equal to $p_\psi$. This is denoted by $\P(\M\satisfies\psi)\geq p_{\psi}$ \cite{BK08}.

\subsection{Barrier certificates}\label{sec:cbc}
Whilst certifying that a stochastic system satisfies a safety specification is generally hard, \emph{barrier certificates} (BC) present a popular tool for safety verification and synthesis.
To this end, we formally define BCs as follows.
\begin{definition}[Barrier certificate (BC)]\label{def:cbc}
	A function $\B:\X\rightarrow\R_{\geq 0}$ is called a BC of an MC $\M=(\X,\X_0,\Tr)$ with reference to an unsafe set $\X_u$, if we have
	\begin{itemize}
		\item[(a)] $\forall x_0\in \X_0:\,\B(x_0)\leq\eta$, 
		\item[(b)] $\forall x_u\in \X_u:\,\B(x_u)\geq\gamma$; and
		\item[(c)] $\forall x\in\X:\,\E_{\Tr}[\B(X^+)|X=x] -\B(x) \leq  c;$
	\end{itemize}
	for some constants $\gamma>\eta\geq0$ and $c\geq 0$.
\end{definition}
Note that condition (c) of Definition~\ref{def:cbc} stipulates a BC exhibiting a relaxed version of the \emph{supermartingale} property for a 
non-zero constant $c$.
The following proposition based on \cite[Theorem~3]{kushner1967stochastic}
establishes quantified guarantees that a system $\M$ satisfies a safety specification $\psi$ if a BC exists.
\begin{proposition}[Finite-horizon safety]\label{prop:reach}
	Consider an MC $\M=(\X,\X_0,\Tr)$ and a safety specification $\psi=(\X_u,T)$. 
	Suppose there exists a BC $B$ w.r.t. $\X_u$ (Definition~\ref{def:cbc}) with constants $(\eta,\gamma,c)$.
	Then, it follows that
	$$\P(\M\satisfies\psi)\geq 1- \frac{\eta + cT}{\gamma}.$$
\end{proposition}
\begin{remark}[Infinite-horizon safety]
	Finding a BC as in Definition~\ref{def:cbc} for a constant $c>0$ is generally easier than for $c=0$. If, however, a BC can be obtained for $c=0$, Proposition~\ref{prop:reach} provides safety guarantees for an unbounded time horizon, i.e., $T\rightarrow\infty$.
\end{remark}

\subsection{Problem statement}
In this paper, we want to verify safety
of a system $\M$ for which the transition kernel $\Tr$ is unknown and only observed through a finite set of observations of the form $\{x^i,x^i_+\}_{i=1}^N$, where $x^i_{+} \sim \Tr(\cdotx|X=x^i)$ for uniformly drawn $x^i\sim\mathcal{U}_\X(\cdotx)$.
This corresponds to drawing i.i.d. samples from the joint distribution $(X,X^+)\sim\Tr(X^+|X)\,\mathcal{U}_\X(X)$.
Note that this setting admits a plethora of target systems.
We formalize the problem statement as follows.

\begin{prob}\label{prob:main}
	\setlength{\belowdisplayskip}{0pt}
	For a given safety specification $\psi$ and confidence $1-\rho\in[0,1]$, find a threshold $p_\psi\in [0,1]$
	without knowledge of the transition kernel $\Tr$ of $\M$ and based only on i.i.d. observations \mbox{$\{x^i,x^i_+\}_{i=1}^N$, $N\in\N_{>0}$,}
	such that
	$\P(\M \satisfies \psi )\!\geq\!   p_\psi$ with probability at least $1-\rho$.
\end{prob}
In this work, we address Problem~\ref{prob:main} under the assumption of some complexity information (see Assumption~\ref{asm:BCinRKHS}) and use BCs to certify probabilistic safety. Indeed, the main challenge in establishing BCs for unknown stochastic systems arises from the chance constraint in condition (c) (Definition~\ref{def:cbc}). We generate BCs based on data by adapting recent advances in RKHS theory and kernel mean embeddings of conditional probability measures to reformulate condition (c) into an inner product with the mean embedding of the unknown transition kernel $\Tr$. We construct an RKHS ambiguity set which is centered at the empirical mean embedding and can be inflated to robustify against the unknown system dynamics.
Based on the reformulated BC constraints, we find a polynomial BC using
an SOS solver and compute its associated RKHS complexity bound by approximating the certificate using Gaussian process (GP) regression.
\smallskip

In the following section, we give a brief introduction to RKHS theory and provide the results for embedding (conditional) probability measures into RKHSs.

\section{RKHS AND KERNEL MEAN EMBEDDINGS}\label{sec:RKHS}
A symmetric function $k_\X:\X\times\X\rightarrow\R$ is called a (positive definite) \emph{kernel} (note the distinction from \emph{probability kernels}) if for all $N\in\N_{>0}$ we have $\sum_{i=1}^{N}\sum_{j=1}^{N}a_ia_jk_\X(x_i,x_j)\geq0$ for $x_1,\ldots,x_N\in\X$ and $a_1,\ldots,a_N\in\R$.
We assume that all kernels in this work are bounded on their domain, i.e., $\E_{}[k_\X(x,x)]<\infty$, $x\in\X$.
Given a kernel $k_\X$ on an non-empty set $\X$, there exists a corresponding unique \emph{reproducing kernel Hilbert space} (RKHS) $\Hilbert_{k_\X}$
of functions $f:\X\rightarrow\R$ equipped with an inner product $\innerH{\cdotx}{\cdotx}{\Hilbert_{k_\X}}$
with the infamous \emph{reproducing property} such that for any function $f\in\Hilbert_{k_\X}$ and $x\in\X$ we have $f(x)=\innerH{f}{k_\X(\cdotx,x)}{\Hilbert_{k_\X}}$.
Note that $k_\X(\cdotx,x):\X\rightarrow\Hilbert_{k_\X}$ is a real-valued function, also called an implicit \emph{canonical} \emph{embedding} or \emph{feature map} $\phi_\X$
such that $k_\X(x,x)=\innerH{\phi_\X(x)}{\phi_\X(x')}{\Hilbert_{k_\X}}$ for all $x,x'\in\X$.
For an RKHS $\Hilbert_{k_\X}$, we use the associated feature map $\phi_\X$ and kernel $k_\X$ interchangeably for ease of notation and comprehensibility.
The inner product induces the norm $\norm{f}_{\Hilbert_{k_\X}}\!\!\!\!:=\!\!\sqrt{\smash[b]{\innerH{f}{f}{\Hilbert_{k_\X}}}}$ of the RKHS. 
A continuous kernel $k_\X$ on a compact metric space $\X$ is \emph{universal} if $\Hilbert_{k_\X}$ is dense in 
continuous functions $C(\X)$ \cite[Sec~4.5]{Steinwart2006SVM}.
Throughout this paper, we assume that all RKHSs are \emph{separable}.
Refer to \cite{Berlinet2004RKHSProbStat} for a comprehensive study on RKHSs.
Given $N$ i.i.d. samples $\hat X:=[\hat{x}_1,\ldots,\hat{x}_N]\T\in\X^N$, the
\emph{Gram matrix} of $k_\X$ is given by
$K_{\hat{X}}:=[k_\X(\hat{x}_i,\hat{x}_j)]_{i,j=1}^N.$
Furthermore, we define the vector-valued function 
$k_{\hat{X}}(x)  := [k_\X(x,\hat{x}_i)]_{i=1}^N .$

\subsection{Embedding probability measures}
To reason about the expected value of a random variable or function, embedding the variable into a (higher dimensional) space is a well-established concept in machine learning \cite{scholkopf2002learning,Steinwart2006SVM}.
The \emph{(kernel) mean embedding} (ME) of a probability measure into Hilbert space is defined as follows \cite{Smola2007EmbedDistrb}.
\begin{definition}[Mean embedding (ME)]
	Given the kernel $k_\X:\X\times\X\rightarrow\R$ spanning an RKHS $\Hilbert_{k_\X}$, 
	the \emph{mean embedding} of a probability measure $p:\borel{\X}\rightarrow[0,1]$
	is given via the \emph{mean map} $\mu_{k_\X}:\mathcal{P}(\X)\rightarrow\Hilbert_{k_\X}$ by
	\begin{equation*}
		\mu_{k_\X}(p) := \E_{p}[\phi_\X(X)] = \int_{\X}\phi_\X(x)\,dp(x).
	\end{equation*}
\end{definition}
Note that the reproducing property carries on to the ME, allowing to compute the expected value of a function $f\in\Hilbert_{k_\X}$
via its inner product with the ME, i.e.,
\begin{equation}
	\E_{p}[f(X)] = \innerH{f}{{\mu}_{k_\X}(p)}{\Hilbert_{k_\X}}.\label{eq:innerProdMean}
\end{equation}
If $k_\X$ is universal, any probability measure $p\in\mathcal{P}(\X)$ is injectively mapped to a unique ME $\mu_{k_\X}(p)\in\Hilbert_{k_\X}$ \cite{Gretton2012}.
This allows the definition of a distance metric between functions in the RKHS. 
For an RKHS $\Hilbert_{k_\X}$ spanned by a universal kernel $k_\X$, such a metric is given by the \emph{maximum mean discrepancy} (MMD). The MMD between two probability measures $p,p'\in\mathcal{P}(\X)$ in $\Hilbert_{k_\X}$ is defined as $\norm{\mu_{k_\X}(p)-\mu_{k_\X}(p')}_{\Hilbert_{k_\X}}$ \cite{Gretton2012}.

\subsection{Embedding conditional probability measures}
Analogous to embedding generic probability measures, we can embed conditional probability measures of the form $\p:\X\times\borel{\Y}\rightarrow[0,1]$ with realizations $Y\sim \p(\cdotx|X=x)$ for a conditioning variable taking values $x\in\X$. 
To this end, we use the measure-theoretic conditional mean embedding (CME) introduced in \cite{Park2020MeasureTheoretic}. 
%
For this, we equip the space of the conditioning random variable $X\in\X$ and the space of the target random variable $Y\in\Y$ with kernels $k_\X:\X\times\X\rightarrow\R$ and $k_\Y:\Y\times\Y\rightarrow\R$, respectively.
\begin{definition}[Conditional mean embedding (CME)]\label{def:condMeanEmbed}
	Given two RKHSs $(\Hilbert_{k_\X},\Hilbert_{k_\Y})$ with the associated kernels $(k_\X,k_\Y)$,
	the CME of a conditional probability measure $\p:\X\times\borel{\Y}\rightarrow[0,1]$
	is an $X$-measurable random variable taking values in $\Hilbert_{k_\Y}$ given by
	\begin{equation*}
		\mu_{k_\Y|k_\X}(\p)(\cdotx) := \E_{\p}[\phi_\Y(Y)\,|\,X=\cdotx].
	\end{equation*}
\end{definition}
Refer to \cite{Park2020MeasureTheoretic} for a comprehensive mathematical dissemination of the CME.
Analogous to the non-conditional case, we can compute the expected value of a function $f\in\Hilbert_{k_\Y}$
via its inner product with the CME, i.e., almost surely
\begin{equation}
	\E_{\p}[f(Y)\,|\,X=x] = \innerH{f}{{\mu}_{k_\Y|k_\X}(\p)(x)}{\Hilbert_{k_\Y}}.\label{eq:innerProdCond}
\end{equation}
Since the CME of a measure $\p$ is generally unknown, we may want to construct an empirical estimate from a finite set of training data.
\begin{proposition}[Empirical CME]\label{prop:empCondMeanEmbed}
	Let a finite dataset $(\hat{X},\hat{Y}):=([\hat{x}_i]_{i=1}^N,[\hat{y}_i]_{i=1}^N) \subset\X\times\Y$ of samples $\hat{y}_i\sim \p(\cdotx|X=\hat{x}_i)$
	from a conditional probability measure $\p:\X\times\borel{\Y}\rightarrow[0,1]$ be given.
	For two kernels $(k_\X,k_\Y)$,
	the \emph{empirical CME} of $\p$ given $(\hat{X},\hat{Y})$ is
	given by
	\begin{equation}
		\hat{\mu}_{k_\Y|k_\X}^N(\cdotx) := k_{\hat{X}}(\cdotx)\T\left[ K_{\hat{X}}+N\lambda I_N\right]^{-1} \phi_\Y(\hat{Y}),\label{eq:empiricalCME}
	\end{equation} 
	with 
	a regularization constant $\lambda\geq0$.
\end{proposition}
Here and in the following, we will assume $[K_{\hat{X}}+N\lambda I_N]$ (and equivalent terms) to be invertible. Note that this is always true if the regularization constant $\lambda$ is strictly positive.
By virtue of the reproducing property we have for any function $f\in\Hilbert_{k_\Y}$ almost surely that 
\begin{equation}
	\innerH{f}{\hat{\mu}^N_{k_\Y|k_\X}(x)}{\Hilbert_{k_\Y}} = k_{\hat{X}}(x)\T\left[ K_{\hat{X}}+N \lambda I_N\right]^{-1} f(\hat{Y}).\label{eq:empiricalInnerProd}
\end{equation}
Note that this yields the conditional expectation of $f(Y)$ given $X=x$ for the empirical conditional probability measure.
\section{DATA-DRIVEN BARRIER CERTIFICATES}\label{sec:ddcbc}

In this section, we establish the supermartingale-like property (c) in Definition~\ref{def:cbc} for a system with unknown transition kernel $\Tr$ using its CME based on i.i.d. training data generated from the unknown true system, i.e., data of the form
\begin{align}\label{eq:data}
	\left. \begin{array}{ll}
		\hat{X}&\!\!:=
		{[\hat{x}_1,\ldots,\hat{x}_N]\T}\\
		\hat{X}^{+\!}&\!\!:=[\hat{x}^+_1,\ldots,\hat{x}^+_N]\T
	\end{array} \right\} \text{where }\hat{x}^+_i\sim\Tr(\cdotx|X=\hat{x}_i).
\end{align}
In the following, we will use two kernels, one for the target variable $x^+$ and one for the conditioning on $x$, namely
\begin{equation}
			k_{+\!}:\X\times\X\rightarrow\R,\quad
			k_{x}:\X\times\X\rightarrow\R,
			\label{eq:kernels}
\end{equation}
with their associated RKHSs $\Hilbert_{k_{+\!}}$ and $\Hilbert_{k_{x}}$, respectively. We will assume that $k_{+\!}$ is universal.
To be able to make statements about the unknown dynamics and establish mathematically rigorous results, we must restrict the system dynamics to a known function space.
In particular, we will assume that the CME of the transition kernel $\Tr$ lives in a vector-valued RKHS of functions from $\X\rightarrow\Hilbert_{k_{+\!}}$, indicated by $\mathcal{G}$ \cite[Sec.~2.3]{Park2020MeasureTheoretic}. We raise the following standard assumption.
\begin{assumption}\label{asm:BCinRKHS}
	Let the CME $\mu_{k_{+\!}|k_{x}}(\Tr)$ be \emph{well-specified}, i.e.,  $\mu_{k_{+\!}|k_{x}}(\Tr)\in\mathcal{G}$.
	Furthermore, for a given confidence \mbox{$1-\rho\in[0,1]$}, let a bound $\varepsilon\geq0$ be known such that $$\P(||\mu_{k_{+\!}|k_{x}}(\Tr)-\hat\mu^N_{k_{+\!}|k_{x}}||_{\mathcal{G}}\leq\varepsilon)\geq 1-\rho,$$ for the empirical CME $\hat\mu^N_{k_{+\!}|k_{x}}$ in \eqref{eq:empiricalCME}.
\end{assumption}
It is common practise to use error bounds such as $\varepsilon$ in Assumption~\ref{asm:BCinRKHS} to robustify results obtained via \eqref{eq:empiricalInnerProd} by constructing an RKHS ambiguity set centered at the empirical CME. There are results available that can relate the radius of the ambiguity set to the probability that the true CME is going to lie within the set with a minimax rate of $\mathcal{O}(\log(N)/N)$~\cite{li2022optimal}.
However, in practice, these concentration bounds are generally excessively conservative, and the radius is commonly chosen manually or using an MMD bootstrap approach (see, e.g., \cite{Nemmour2022FiniteSampleGuarantee} for non-conditional mean embeddings).

To establish barrier certificates, we start by reformulating the left-hand side of condition (c) in Definition~\ref{def:cbc} as an inner product with the CME $\mu_{k_{+\!}|k_{x}}(\Tr)$ via \eqref{eq:innerProdCond}, i.e., we have
\begin{equation*}
	\E_{\Tr}[\B(X^+)|X=x] = \innerH{\B}{\mu_{k_{+\!}|k_{x}}(\Tr)(x)}{\Hilbert_{k_{+\!}}}.
\end{equation*}
Since the CME $\mu_{k_{+\!}|k_{x}}(\Tr)$ is unknown, we construct an ambiguity set $\ambRKHS^{N}_\varepsilon\subset\mathcal{G}$ centered at the empirical CME $\hat\mu^N_{k_{+\!}|k_{x}}$ constructed from the data in \eqref{eq:data} (via Proposition~\ref{prop:empCondMeanEmbed}) and choose an MMD radius $\varepsilon\geq0$ such that the CME of $\Tr$ lies within $\ambRKHS^{N}_\varepsilon$ with a confidence of at least $1-\rho\in[0,1]$, i.e., such that $\P(\mu_{k_{+\!}|k_{x}}(\Tr)\in\ambRKHS^{N}_\varepsilon)\geq 1-\rho
	$, where
\begin{equation}\label{eq:ambRKHS}
	\textstyle\!\!\!\!\!\ambRKHS^{N}_\varepsilon
	\! := \! \left\lbrace\mu\in\mathcal{G} \left|\, \norm{\mu-\hat\mu^N_{k_{+\!}|k_{x}}}_{\mathcal{G}} \leq \varepsilon\right.\right\rbrace.
\end{equation}
The next result follows trivially from \eqref{eq:innerProdCond}.
\begin{theorem}\label{thm:cbc1}
	Let data $(\hat{X},\hat{X}^+)$ in \eqref{eq:data} from an unknown MC $\M$ and
	kernels $(k_{+\!},k_{x})$ in \eqref{eq:kernels} ($k_{+\!}$ universal) be given.
	Consider the empirical CME $\hat\mu^N_{k_{+\!}|k_{x}}$ given $(k_{+\!},k_{x})$ and $(\hat{X},\hat{X}^+)$, and the ambiguity set $\ambRKHS^{N}_\varepsilon$ in \eqref{eq:ambRKHS} with confidence $1-\rho$.
	If there exists	a function $\B:\X\rightarrow\R_{\geq 0}$ satisfying
	\begin{equation}
		\forall x\in\X,\,\forall \mu\in\ambRKHS^{N}_\varepsilon: \, \innerH{\B}{\mu(x)}{\Hilbert_{k_{+\!}}} -\B(x)  \leq c,\label{eq:ineqCondForallX}
	\end{equation}
	for some constant $c\geq 0$,
	then
	$\B$ satisfies BC condition (c) of Definition~\ref{def:cbc} w.r.t. $\M$ with probability at least $1-\rho$.
\end{theorem}

With the following theorem, we provide a way of establishing Theorem~\ref{thm:cbc1} by computing the left-hand side of \eqref{eq:ineqCondForallX} for the RKHS norm-ball ambiguity set in \eqref{eq:ambRKHS}.
\begin{theorem}\label{thm:cbc2}
	Let data $(\hat{X},\hat{X}^+)$ in \eqref{eq:data} from an unknown MC $\M$, 
	kernels $(k_{+\!},k_{x})$ in \eqref{eq:kernels} ($k_{+\!}$ universal) with the corresponding Gram matrices $K_{+\!}$ and $K_{{\hat{X}}}$ based on $\hat{X}^+$ and $\hat{X}$, respectively, be given.
	Consider the ambiguity set $\ambRKHS^{N}_\varepsilon$ in \eqref{eq:ambRKHS} with $1-\rho$.
	If there exists	a function $\B:\X\rightarrow\R_{\geq 0}$ satisfying $B\in\Hilbert_{k_{+\!}}$ with some $\bar{B}\geq\norm{B}_{\Hilbert_{k_{+\!}}}$ and for which
	\begin{equation}
		\forall x\in\X:\,
		w(x)\T
		\B(\hat{X}^+) -\B(x)\leq  c - \varepsilon\sqrt{k_x(x,x)}
		\bar{B}
		,\label{eq:ineqCondForallX2}
	\end{equation}
	for some constant $c\geq 0$, and 
	\begin{equation}
		w(x)\T:=k_{\hat{X}}(x)\T\left[ K_{\hat{X}}+N\lambda I_N\right]^{-1},\label{eq:weightfcn}\\
	\end{equation}
	with constant $\lambda\geq 0$
	and function $k_{\hat{X}}(x) := [k_x(x,\hat{x}_i)]_{i=1}^N$, 
	then $\B$ satisfies BC condition (c) of Definition~\ref{def:cbc} w.r.t. $\M$ with probability at least $1-\rho$.
\end{theorem}
\begin{proof}
	We start by rewriting the inner product in \eqref{eq:ineqCondForallX} as
	\begin{equation*}
		\innerH{\B}{\mu(x)}{\Hilbert_{k_{+\!}}} \!\!=\! 
		\innerH{\B}{\mu(x)\!-\!\hat\mu^N_{k_{+\!}|k_{x}}\!(x)}{\Hilbert_{k_{+\!}}} \!\!+\! \innerH{\B}{\hat\mu^N_{k_{+\!}|k_{x}}\!(x)}{\Hilbert_{k_{+\!}}}. 
	\end{equation*}
	Via \eqref{eq:empiricalInnerProd}, the latter term yields	$w(x)\T\B(\hat{X}^+)$.
	For the prior term, let $\Gamma:\X\times\X\rightarrow\mathcal{L}(\Hilbert_{k_{+\!}})$ be the operator-valued positive definite kernel of $\mathcal{G}$ (see, e.g., \cite[Def.~1]{li2022optimal}) given by $\Gamma(x,x'):=k_x(x,x')\mathrm{Id}_{\Hilbert_{k_{+\!}}}$, where $\mathcal{L}(\Hilbert_{k_{+\!}})$ is the Banach space of bounded linear operators from $\Hilbert_{k_{+\!}}$ to $\Hilbert_{k_{+\!}}$ and $\mathrm{Id}_{\Hilbert_{k_{+\!}}}$ is the identity operator on $\Hilbert_{k_{+\!}}$.
	Then, it follows from the reproducing property of $\Gamma$ (cf. \cite[Equation~(2)ff.]{li2022optimal}) that
	\begin{align*}
		\innerH{\B}{\mu(x)\!-\!\hat\mu^N_{k_{+\!}|k_{x}}\!(x)}{\Hilbert_{k_{+\!}}} &= \innerH{\Gamma(\cdotx,x)\B}{\mu-\hat\mu^N_{k_{+\!}|k_{x}}}{\mathcal{G}},\\
		&\leq \varepsilon \norm{\Gamma(\cdotx,x)\B}_{\mathcal{G}},\\
		&= \varepsilon \sqrt{\innerH{\Gamma(\cdotx,x)\B}{\Gamma(\cdotx,x)\B}{\mathcal{G}}},\\
		&= \varepsilon \sqrt{\innerH{\B}{\Gamma(x,x)\B}{\Hilbert_{k_{+\!}}}},\\
		&\leq \varepsilon\bar{B}\sqrt{\norm{\Gamma(x,x)}},\\
		&= \varepsilon\bar{B}\sqrt{k_x(x,x)}.
	\end{align*}
	Reordering yields \eqref{eq:ineqCondForallX2}, concluding the proof. 
\end{proof}
Intuitively, the conservatism introduced
due to the worst-case approach w.r.t. the unknown transition kernel $\Tr$
is captured by the term $\varepsilon\sqrt{k_x(x,x)}\bar{B}$, hence proportional to the radius $\varepsilon$. Ensuring that \eqref{eq:ineqCondForallX2} holds for an adequate offset $\varepsilon\sqrt{k_x(x,x)}\bar{B}$ gives the necessary headroom to provide guarantees for transition kernels deviating from the empirical CME.
Note that, as we design the BC $B$, an upper bound $\bar{B}$ can be estimated via either analytical expressions or numerical estimation \cite{Berlinet2004RKHSProbStat,Kanagawa2018GPvsKernel}.
Based on Theorem~\ref{thm:cbc2}, we provide results on system safety via Proposition~\ref{prop:reach}.
\begin{theorem}[Data-driven finite-horizon safety]\label{thm:safety}
	Consider the setup in Theorem~\ref{thm:cbc2} and a safety specification $\psi=(\X_u,T)$.
	Suppose there exists a function $B$ 
	satisfying the conditions in Theorem~\ref{thm:cbc2}
	for a constant $c\geq 0$.
	If there exist constants $\gamma>\eta\geq0$ such that
	\begin{itemize}
		\item[(a)] $\forall x_0\in \X_0:\,\B(x_0)\leq\eta$; and
		\item[(b)] $\forall x_u\in \X_u:\,\B(x_u)\geq\gamma$; 
	\end{itemize}
	are satisfied, then, with probability at least $1-\rho$ we have
	$$\P(\M\satisfies\psi)\geq 1- \frac{\eta + cT}{\gamma} .$$
\end{theorem}
Theorem~\ref{thm:safety} follows trivially from Theorem~\ref{thm:cbc2} and Proposition~\ref{prop:reach}.

\section{SUM-OF-SQUARES FORMULATION}\label{sec:sos}
Scrutinizing general functions against the conditions in Theorems~\ref{thm:cbc2} and \ref{thm:safety} to obtain a valid BC is generally challenging. Hence, restricting the class of functions for, e.g., the barrier function $\B$, is a common approach.
We make the following choices for $\B$, $k_{x}$, and $k_{+\!}$, to obtain a program that can be efficiently solved using SOS solvers \cite{Parrilo2003SOS}.
We let $\B$ be a polynomial function of $x\in\X\subset\R^n$ and linear w.r.t. parameters $(b_0,\ldots,b_q)\in\R^q$, e.g., for $n=2$ and $q=5$ we have
$\B(x) := b_0 + b_1x_1 + b_2x_2 + b_3x_1^2 + b_4x_1x_2 + b_5x_2^2$.
We denote the space of polynomial functions as $\Hilbert_{\text{poly}}$.
For $k_{x}$ we pick the polynomial kernel of degree $d\in\N_{>0}$, i.e., $$k_{x}(x,x'):=\left(a(x\T x')+b\right)^d,\quad a>0,\,b\geq 0,$$ rendering $w(x)\T$ in \eqref{eq:weightfcn} polynomial in $x$.
%
%
Furthermore, we assume that the state space $\X$ is continuous
and the sets $(\X,\X_0,\X_u)$ are given in semi-algebraic form as $\X:=\{x:\,g_X(x)\geq 0\}$, $\X_0:=\{x\in\X:\,g_{0}(x)\geq 0\}$, and $\X_u:=\{x\in\X:\,g_{u}(x)\geq 0\}$, where $g_X(x)$, $g_0(x)$, and $g_u(x)$ are vector-valued functions.
For the universal kernel $k_{+\!}$, we select the squared-exponential kernel \cite{Kanagawa2018GPvsKernel, Rasmussen2005GP}, i.e.,
\begin{equation*}
	k_{+\!}(x, x') := \sigma_f^2\exp\left\lbrace -\dfrac{1}{2 \sigma_l^2}\norm{x-x'}^2 \right\rbrace,
\end{equation*}
with hyperparameters $\sigma_f^2,\sigma_l^2>0$.
Although the squared-exponential kernel is universal and therefore dense in continuous functions, it does not contain any polynomial functions \cite[Theorem~2]{Minh2010Gaussian}.
Hence, it follows that the polynomial barrier $B\notin\Hilbert_{k_{+\!}}$ and the function $\tilde\B\in\Hilbert_{k_{+\!}}$ approximating $\B\in\Hilbert_{\text{poly}}$ \emph{exactly} is infinitely complex, i.e.,  $||\tilde\B||_{\Hilbert_{k_{+\!}}}\rightarrow\infty$ \cite[Sec.~1.2]{Dette2021RKHSPoly}.
We hence propose the following three-step approach.

\smallskip

\noindent\textbf{1) Compute conservative barrier via SOS:}
To obtain an approximation $\tilde\B\in\Hilbert_{k_{+\!}}$ with finite RKHS norm, i.e., $||\tilde\B||_{\Hilbert_{k_{+\!}}}\leq\bar{B}$ with some $0\leq\bar{B}<\infty$, we generate a polynomial $\B\in\Hilbert_{\text{poly}}$ that satisfies a slightly more conservative version of Theorem~\ref{thm:safety}, i.e., for all $x\in\X$
\begin{align}
	\begin{split}
	\B(x)-\zeta_1 -\lambda_X(x)\T g_X(x) &\geq 0 
	,\\
	\gamma+\zeta_1 - \B(x) -\lambda_0(x)\T g_0(x) &\geq 0 
	,\\
	\B(x)-\zeta_1 -\eta -\lambda_u(x)\T g_u(x) &\geq 0 
	,\\
	\B(x)  -w(x)\T\B(\hat{X}^+) -\xi-\lambda_X(x)\T g_X(x) &\geq 0,
	\end{split}\label{eq:SOS}
\end{align}
with $\xi:=(\varepsilon\sup_{x\in\X}\sqrt{k_x(x,x)}\bar{B}-c + \zeta_1+ \zeta_2)$,
slack variables $\zeta_1,\zeta_2> 0$, and vector-valued functions of positive polynomials $\lambda_X(x),\lambda_0(x),\lambda_u(x)\geq 0$.
This problem can be solved as an SOS program via a tool such as SOSTOOLS \cite{sostools}.
Refer to the paper \cite[Sec.~V.C.1]{Pushpak2021CBC} for a more detailed dissemination of the SOS approach.

\smallskip

\noindent\textbf{2) Compute approximation via GP regression:}
We draw $\tilde{N}$ samples $\tilde{y}_i = \B(\tilde{x}_i)$, $(\tilde{x}_1,\ldots,\tilde{x}_{\tilde{N}})\subset\X$ to compute an approximation $\tilde\B\in\Hilbert_{k_{+\!}}$ of $\B$ using Gaussian process regression \cite{Rasmussen2005GP} s.t. $\sup_{x\in\X}|\B(x)-\tilde{\B}(x)|\leq\zeta_1$ and $\sup_{x\in\X} |w(x)\T(\B(\hat{X}^+)-\tilde{\B}(\hat{X}^+))|\leq\zeta_2$.

\smallskip

\noindent\textbf{3) Compute RKHS norm:} The RKHS norm of $\tilde{\B}$ is obtained via $||\tilde\B||_{\Hilbert_{k_{+\!}}}=\sqrt{\alpha\T K\alpha}$ with $\alpha:=[K+\lambda I]^{-1}[\tilde\B(\tilde{x}_i)]_{i=1}^{\tilde{N}}$ \cite{Romao2023DRControl}, for which we require $||\tilde\B||_{\Hilbert_{k_{+\!}}}\leq\bar{B}$ to hold.

\smallskip

Note that the approach above might require iterating over several instances of $(\bar{B},\zeta_1,\zeta_2)$.

\section{NUMERICAL RESULTS}\label{sec:numerical_studies}
\begin{figure}
	\centering
	\includegraphics[width=1\columnwidth]{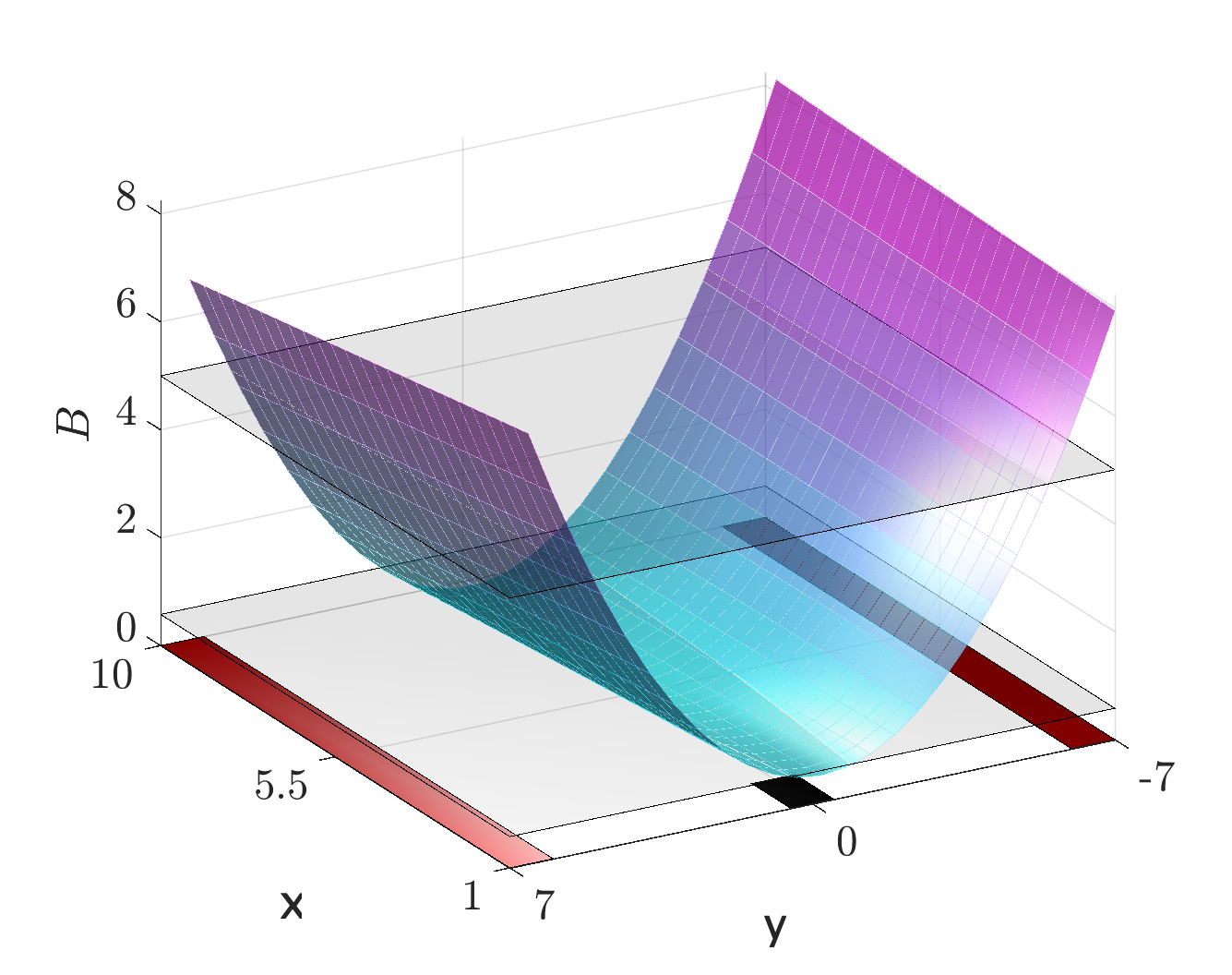}
	\caption{Polynomial barrier certificate computed using SOSTOOLS depicted for $\upphi=0$. The unsafe regions $\X_u$ are indicated in red and the initial region $\X_0$ in black. The two gray planes represent the level sets $\eta$ and $\gamma$.}
	\label{fig:CSLaneKeeping}
\end{figure}

To put the presented approach into context with the current state-of-the-art, we demonstrate its performance on the lane keeping case study from {Salamati et al.} \cite{Salamati2021DDCBC}, adapted from the single-track model from the book \cite[Sec.~2.2]{rajamani2011vehicle}.
%
To this end, consider a car governed by the following dynamics
\begin{align}
	\begin{bmatrix} \mathsf{x}_{t+1}\\\mathsf{y}_{t+1}\\\upphi_{t+1} \end{bmatrix} = \begin{bmatrix} \mathsf{x}_{t}\\\mathsf{y}_{t}\\\upphi_{t} \end{bmatrix} + 
	\begin{bmatrix} 
		\tau v\cos(\upphi_t + \uppsi)\\
		\tau v \sin(\upphi_t + \uppsi)\\
		\tau \frac{v}{l_r} \sin(\uppsi)
	\end{bmatrix} + \begin{bmatrix} w^1_t\\w^2_t\\w^3_t  \end{bmatrix},\label{eq:car}
\end{align}
where $\mathsf{x}$, $\mathsf{y}$, $\upphi$ denote the position in longitudinal direction, lateral direction, and the heading angle, respectively.
The goal is to compute the probability that the system initialized in $[\mathsf{x}_0,\mathsf{y}_0,\upphi_0]\T\in \X_0$ does not enter the unsafe region $\X_u$ within $T=10$ time steps (cf. Fig.~\ref{fig:CSLaneKeeping}).
The noise $w_t^i$, $i\in\{1,2,3\}$, is zero-mean Gaussian with standard deviation $0.01$, $0.01$, and $0.001$, respectively.
We use the parameters $(\tau=0.1, v=5, l_r=l_f=1.384, \uppsi=\frac{l_r}{l_r+l_f}\arctan(\delta_f),\delta_f=5^\circ)$ and sets $\X=[1,10]\times[-7,7]\times[-0.05,0.05]$, $\X_0=[1,2]\times[-0.5,0.5]\times[-0.005,0.005]$, $\X_u=\X_{u1}\cup\X_{u2}$ with $\X_{u1}=[1,10]\times[-7,-6]\times[-0.05,0.05]$ and $\X_{u2}=[1,10]\times[6,7]\times[-0.05,0.05]$ (cf. Fig.~\ref{fig:CSLaneKeeping}) as in \cite{Salamati2021DDCBC}.

\begin{figure}
	\centering
	\includegraphics[width=\columnwidth]{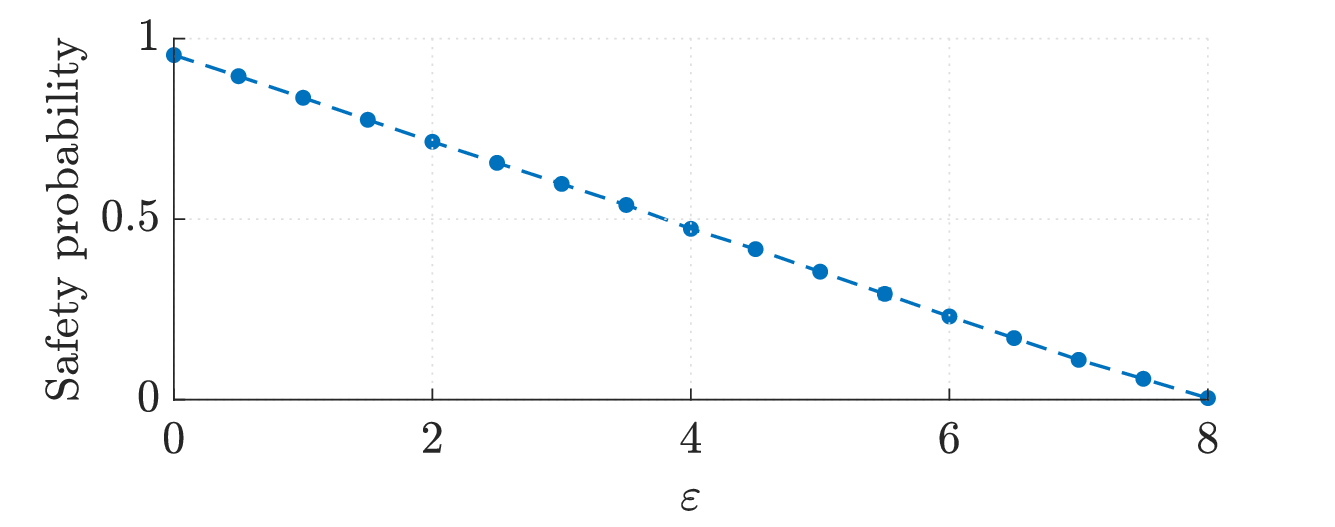}
	\caption{Robust safety probability via BCs generated for radii $\varepsilon$, averaged over ten samples of size $N=10^4$ and a constant $\bar{B}=0.1$.}
	\label{fig:probOverRadius}
\end{figure}

For the polynomial and squared-exponential kernels $k_{x}$ and $k_{+\!}$, we select the hyperparameters $(a=0.005,b=0.11, d=2)$ and $(\sigma_f^2=1500^2,\sigma_l=2.98^2)$, respectively.
We draw $N=10^4$ samples from the system in \eqref{eq:car}.
For the computation of \eqref{eq:weightfcn}, the constant $\lambda=10^{-3}$ is used.
We solve the SOS program in \eqref{eq:SOS} for varying radii $\varepsilon$ using SOSTOOLS \cite{sostools}. For $\varepsilon=1$ we find a BC using $(c=10^{-4},\bar{B}=0.06,\zeta_1=0.01,\zeta_2=0.01)$, namely
\begin{align*}
	&\B(x)= 
	-1.425\cdot10^{-4}\mathsf{x}^2
	- 0.012\mathsf{x}\mathsf{y}
	- 0.028\mathsf{x}\upphi
	+ 0.162\mathsf{y}^2\\
	&\!+\! 0.774\mathsf{y}\upphi
	+\! 0.716x\upphi^2\!
	- \!0.048\mathsf{x}
	+\! 0.044\mathsf{y}
	-\! 0.270\upphi
	+\! 0.562
\end{align*}
 depicted in Fig.~\ref{fig:CSLaneKeeping},
with level sets $(\gamma=5,\eta=0.58)$, certifying that the system remains safe for $T$ time steps with a probability of at least $88\%$ (Theorem~\ref{thm:safety}).
We approximate $\B$ via GP regression s.t. the error bounds $(\zeta_1,\zeta_2)$ are satisfied (Sec.~\ref{sec:sos}).
For the approximation $\tilde{\B}$, we verify that $||\tilde\B||_{\Hilbert_{k_{+\!}}}\leq\bar{B}$.
We repeat the steps mentioned for $\bar{B}=0.1$ and cascading radii $\varepsilon$ and report the corresponding robust safety probabilities in Fig.~\ref{fig:probOverRadius}.
Note that we certify safety requiring minimal regularity assumptions. In contrast, \cite{Salamati2021DDCBC} assumes knowledge of the Lipschitz constants of several intermediate functions and every system state must be sampled 3000 times.

\section{CONCLUDING REMARKS}\label{sec:conclusion}
In future work, we plan to extend the presented approach to control synthesis and complex specifications beyond safety. This includes languages such as LTL over finite traces (LTL\textsubscript{F}) \cite{Pushpak2021CBC} as well as accommodating multiple specifications \cite{Wang2016aMultiBC}.
Further investigation is needed to determine sharp ambiguity set radii $\varepsilon$ and reduce the conservatism of the approach, including alternative concentration theorems tailored to the class of barrier functions (e.g., via Rademacher complexity) and alternative algorithmic solutions to Theorem~\ref{thm:cbc1}.
A computational bottleneck of the presented approach is the computation of the inverse of an $N\times N$ kernel matrix.


\bibliographystyle{abbrv}
\bibliography{references}

\begin{thebibliography}{10}

\bibitem{BK08}
C.~Baier and J.-P. Katoen.
\newblock {\em Principles of Model Checking}.
\newblock MIT Press, 2008.

\bibitem{belta2017formal}
C.~Belta, B.~Yordanov, and E.~A. Gol.
\newblock {\em Formal methods for discrete-time dynamical systems}, volume~15.
\newblock Springer, 2017.

\bibitem{Berlinet2004RKHSProbStat}
A.~Berlinet and C.~Thomas-Agnan.
\newblock {\em Reproducing Kernel Hilbert Spaces in Probability and
  Statistics}.
\newblock Kluwer Academic, Boston, 2004.

\bibitem{bogachev2007measure}
V.~I. Bogachev.
\newblock {\em Measure theory}.
\newblock Springer Science \& Business Media, 2007.

\bibitem{Chekan2023UncertainConstraints}
J.~A. Chekan and C.~Langbort.
\newblock Safety-aware learning-based control of systems with uncertainty
  dependent constraints.
\newblock In {\em 2023 American Control Conference (ACC)}, pages 1264--1270,
  2023.

\bibitem{Cohen2022}
M.~H. Cohen, C.~Belta, and R.~Tron.
\newblock Robust control barrier functions for nonlinear control systems with
  uncertainty: A duality-based approach.
\newblock In {\em Proceedings of the IEEE Conference on Decision and Control},
  volume 2022-Decem, pages 174--179, aug 2022.

\bibitem{Dette2021RKHSPoly}
H.~Dette and A.~A. Zhigljavsky.
\newblock Reproducing kernel {Hilbert} spaces, polynomials, and the classical
  moment problem.
\newblock {\em SIAM-ASA Journal on Uncertainty Quantification},
  9(4):1589--1614, 2021.

\bibitem{gracia2023distributionally}
I.~Gracia, D.~Boskos, L.~Laurenti, and M.~Mazo~Jr.
\newblock Distributionally robust strategy synthesis for switched stochastic
  systems.
\newblock In {\em Proceedings of the 26th ACM International Conference on
  Hybrid Systems: Computation and Control}, pages 1--10, 2023.

\bibitem{Gretton2012}
A.~Gretton, K.~M. Borgwardt, M.~J. Rasch, B.~Sch{\"{o}}lkopf, and A.~Smola.
\newblock A kernel two-sample test.
\newblock {\em Journal of Machine Learning Research}, 13(25):723--773, 2012.

\bibitem{jackson2020safety}
J.~Jackson, L.~Laurenti, E.~Frew, and M.~Lahijanian.
\newblock Safety verification of unknown dynamical systems via {Gaussian}
  process regression.
\newblock In {\em 2020 59th IEEE Conference on Decision and Control (CDC)},
  pages 860--866. IEEE, 2020.

\bibitem{Jagtap2020CBCGP}
P.~Jagtap, G.~J. Pappas, and M.~Zamani.
\newblock Control barrier functions for unknown nonlinear systems using
  {Gaussian} processes.
\newblock In {\em Proceedings of the IEEE Conference on Decision and Control},
  volume 2020-Decem, pages 3699--3704, oct 2020.

\bibitem{Pushpak2021CBC}
P.~Jagtap, S.~Soudjani, and M.~Zamani.
\newblock Formal synthesis of stochastic systems via control barrier
  certificates.
\newblock {\em IEEE Transactions on Automatic Control}, 66(7):3097--3110, 2021.

\bibitem{Kanagawa2018GPvsKernel}
M.~Kanagawa, P.~Hennig, D.~Sejdinovic, and B.~K. Sriperumbudur.
\newblock Gaussian processes and kernel methods: A review on connections and
  equivalences.
\newblock {\em arXiv:1807.02582}, 2018.

\bibitem{kazemi2022datadriven}
M.~Kazemi, R.~Majumdar, M.~Salamati, S.~Soudjani, and B.~Wooding.
\newblock Data-driven abstraction-based control synthesis.
\newblock {\em Nonlinear Analysis: Hybrid Systems}, 52:101467, 2024.

\bibitem{kazemi2020fullLTL}
M.~Kazemi and S.~Soudjani.
\newblock Formal policy synthesis for continuous-state systems via
  reinforcement learning.
\newblock In {\em Integrated Formal Methods: 16th International Conference, IFM
  2020, Lugano, Switzerland, November 16--20, 2020, Proceedings 16}, pages
  3--21. Springer, 2020.

\bibitem{Klebanov2020RigorousCME}
I.~Klebanov, I.~Schuster, and T.~J. Sullivan.
\newblock A rigorous theory of conditional mean embeddings.
\newblock {\em SIAM Journal on Mathematics of Data Science}, 2(3):583--606,
  2020.

\bibitem{kushner1967stochastic}
H.~J. Kushner and Kushner.
\newblock {\em Stochastic stability and control}, volume~33.
\newblock Academic press New York, 1967.

\bibitem{lavaei2020RL}
A.~Lavaei, F.~Somenzi, S.~Soudjani, A.~Trivedi, and M.~Zamani.
\newblock Formal controller synthesis for continuous-space {MDPs} via
  model-free reinforcement learning.
\newblock In {\em 2020 ACM/IEEE 11th International Conference on Cyber-Physical
  Systems (ICCPS)}, pages 98--107, 2020.

\bibitem{Lavaei_Survey}
A.~Lavaei, S.~Soudjani, A.~Abate, and M.~Zamani.
\newblock Automated verification and synthesis of stochastic hybrid systems: A
  survey.
\newblock {\em Automatica}, 146:110617, 2022.

\bibitem{li2022optimal}
Z.~Li, D.~Meunier, M.~Mollenhauer, and A.~Gretton.
\newblock Optimal rates for regularized conditional mean embedding learning.
\newblock {\em Advances in Neural Information Processing Systems},
  35:4433--4445, 2022.

\bibitem{lindemann2021learning}
L.~Lindemann, H.~Hu, A.~Robey, H.~Zhang, D.~Dimarogonas, S.~Tu, and N.~Matni.
\newblock Learning hybrid control barrier functions from data.
\newblock In {\em Conference on Robot Learning}, pages 1351--1370. PMLR, 2021.

\bibitem{Lopez2022uCBF}
B.~T. Lopez and J.-J.~E. Slotine.
\newblock Unmatched control barrier functions: Certainty equivalence adaptive
  safety.
\newblock In {\em 2023 American Control Conference (ACC)}, pages 3662--3668,
  2023.

\bibitem{makdesi2023data}
A.~Makdesi, A.~Girard, and L.~Fribourg.
\newblock Data-driven models of monotone systems.
\newblock {\em IEEE Transactions on Automatic Control}, 2023.

\bibitem{Minh2010Gaussian}
H.~Q. Minh.
\newblock Some properties of {Gaussian} reproducing kernel {Hilbert} spaces and
  their implications for function approximation and learning theory.
\newblock {\em Constructive Approximation}, 32(2):307--338, dec 2010.

\bibitem{Nemmour2022FiniteSampleGuarantee}
Y.~Nemmour, H.~Kremer, B.~Schölkopf, and J.-J. Zhu.
\newblock Maximum mean discrepancy distributionally robust nonlinear
  chance-constrained optimization with finite-sample guarantee.
\newblock In {\em 2022 IEEE 61st Conference on Decision and Control (CDC)},
  pages 5660--5667, 2022.

\bibitem{sostools}
A.~Papachristodoulou, J.~Anderson, G.~Valmorbida, S.~Prajna, P.~Seiler, and
  P.~A. Parrilo.
\newblock {\em {SOSTOOLS}: Sum of squares optimization toolbox for {MATLAB}}.
\newblock \texttt{http://arxiv.org/abs/1310.4716}, 2013.
\newblock Available from \texttt{http://www.eng.ox.ac.uk/control/sostools}.

\bibitem{Park2020MeasureTheoretic}
J.~Park and K.~Muandet.
\newblock A measure-theoretic approach to kernel conditional mean embeddings.
\newblock In {\em Advances in Neural Information Processing Systems}, volume
  2020-Decem, pages 21247--21259, 2020.

\bibitem{Parrilo2003SOS}
P.~A. Parrilo.
\newblock Semidefinite programming relaxations for semialgebraic problems.
\newblock {\em Mathematical Programming, Series B}, 96(2):293--320, may 2003.

\bibitem{prajna2006barrier}
S.~Prajna.
\newblock Barrier certificates for nonlinear model validation.
\newblock {\em Automatica}, 42(1):117--126, 2006.

\bibitem{rajamani2011vehicle}
R.~Rajamani.
\newblock {\em Vehicle dynamics and control}.
\newblock Springer Science \& Business Media, 2011.

\bibitem{Rasmussen2005GP}
C.~E. Rasmussen and C.~K.~I. Williams.
\newblock {\em Gaussian Processes for Machine Learning}.
\newblock The MIT Press, 11 2005.

\bibitem{Romao2023DRControl}
L.~Romao, A.~R. Hota, and A.~Abate.
\newblock Distributionally robust optimal and safe control of stochastic
  systems via kernel conditional mean embedding.
\newblock In {\em 2023 62nd IEEE Conference on Decision and Control (CDC)},
  pages 2016--2021, 2023.

\bibitem{Salamati2021DDCBC}
A.~Salamati, A.~Lavaei, S.~Soudjani, and M.~Zamani.
\newblock Data-driven verification and synthesis of stochastic systems via
  barrier certificates.
\newblock {\em Automatica}, 159:111323, 2024.

\bibitem{scholkopf2002learning}
B.~Sch{\"o}lkopf and A.~J. Smola.
\newblock {\em Learning with kernels: Support vector machines, regularization,
  optimization, and beyond}.
\newblock MIT press, 2002.

\bibitem{schoen2023bayesian}
O.~Schön, B.~van Huijgevoort, S.~Haesaert, and S.~Soudjani.
\newblock Bayesian formal synthesis of unknown systems via robust simulation
  relations.
\newblock {\em arXiv:2304.07428}, 2023.

\bibitem{Smola2007EmbedDistrb}
A.~Smola, A.~Gretton, L.~Song, and B.~Sch{\"{o}}lkopf.
\newblock A {Hilbert} space embedding for distributions.
\newblock In {\em Lecture Notes in Computer Science (including subseries
  Lecture Notes in Artificial Intelligence and Lecture Notes in
  Bioinformatics)}, volume 4754 LNAI, pages 13--31, 2007.

\bibitem{Song2009CondEmbed}
L.~Song, J.~Huang, A.~Smola, and K.~Fukumizu.
\newblock Hilbert space embeddings of conditional distributions with
  applications to dynamical systems.
\newblock In {\em Proceedings of the 26th Annual International Conference on
  Machine Learning}, ICML '09, page 961–968, New York, NY, USA, 2009.
  Association for Computing Machinery.

\bibitem{Steinwart2006SVM}
I.~Steinwart and A.~Christmann.
\newblock {\em {Support Vector Machines}}.
\newblock Information Science and Statistics. Springer New York, New York, NY,
  2008.

\bibitem{Wang2016aMultiBC}
L.~Wang, A.~D. Ames, and M.~Egerstedt.
\newblock Multi-objective compositions for collision-free connectivity
  maintenance in teams of mobile robots.
\newblock In {\em 2016 IEEE 55th Conference on Decision and Control, CDC 2016},
  pages 2659--2664, aug 2016.

\bibitem{Wang2018CBF}
L.~Wang, E.~A. Theodorou, and M.~Egerstedt.
\newblock Safe learning of quadrotor dynamics using barrier certificates.
\newblock In {\em Proceedings - IEEE International Conference on Robotics and
  Automation}, pages 2460--2465, oct 2018.

\bibitem{yin2024formal}
X.~Yin, B.~Gao, and X.~Yu.
\newblock Formal synthesis of controllers for safety-critical autonomous
  systems: Developments and challenges.
\newblock {\em arXiv:2402.13075}, 2024.

\bibitem{zhang2024formal}
Z.~Zhang, C.~Ma, S.~Soudijani, and S.~Soudjani.
\newblock Formal verification of unknown stochastic systems via non-parametric
  estimation.
\newblock {\em International Conference on Artificial Intelligence and
  Statistics (AISTATS)}, 2024.

\end{thebibliography}
\end{document}